\documentclass[preprint2]{aastex63}

\usepackage{graphicx,epsfig,fancyhdr,psfig,rotating,amsmath,epsf,txfonts,natbib,epstopdf,multirow,booktabs,longtable,color}

\received{***}
\revised{***}
\accepted{19 August 2021}
\submitjournal{ApJL}

\shorttitle{EUI coronal microjets}
\shortauthors{Hou et al.}

\def \kms {{\rm km\;s$^{-1}$}}

\def \hi {H\,{\sc i}}
\def \lya {Ly$\alpha$}
\graphicspath{{./}}

\begin{document}
\title{Coronal microjets in quiet-Sun regions observed with the Extreme Ultraviolet Imager onboard Solar Orbiter}

\correspondingauthor{Hui Tian}
\email{huitian@pku.edu.cn}

\author{Zhenyong Hou}
\affiliation{School of Earth and Space Sciences, Peking University, Beijing, 100871, China}

\author{Hui Tian}
\affiliation{School of Earth and Space Sciences, Peking University, Beijing, 100871, China}
\affiliation{Key Laboratory of Solar Activity, National Astronomical Observatories, Chinese Academy of Sciences, Beijing, \\
100012, China}

\author{David Berghmans}
\affiliation{Solar-Terrestrial Centre of Excellence $-$ SIDC, Royal Observatory of Belgium, Ringlaan -3- Av. Circulaire, 1180 Brussels, Belgium}

\author{Hechao Chen}
\affiliation{School of Earth and Space Sciences, Peking University, Beijing, 100871, China}

\author{Luca Teriaca}
\affiliation{Max-Planck-Institut f\"ur Sonnensystemforschung, G\"ottingen, Germany}

\author{Udo Sch\"uhle}
\affiliation{Max-Planck-Institut f\"ur Sonnensystemforschung, G\"ottingen, Germany}

\author{Yuhang Gao}
\affiliation{School of Earth and Space Sciences, Peking University, Beijing, 100871, China}

\author{Yajie Chen}
\affiliation{School of Earth and Space Sciences, Peking University, Beijing, 100871, China}

\author{Jiansen He}
\affiliation{School of Earth and Space Sciences, Peking University, Beijing, 100871, China}

\author{Linghua Wang}
\affiliation{School of Earth and Space Sciences, Peking University, Beijing, 100871, China}

\author{Xianyong Bai}
\affiliation{Key Laboratory of Solar Activity, National Astronomical Observatories, Chinese Academy of Sciences, Beijing, \\
100012, China}

\begin{abstract}
We report the smallest coronal jets ever observed in the quiet Sun with recent high-resolution observations from the High Resolution Telescopes (HRI$_{EUV}$ and HRI$_{Ly\alpha}$) of the Extreme Ultraviolet Imager (EUI) onboard Solar Orbiter. In the HRI$_{EUV}$ (174 \AA) images, these microjets usually appear as nearly collimated structures with brightenings at their footpoints. Their average lifetime, projected speed, width, and maximum length are 4.6 min, 62 km s$^{-1}$, 1.0\,Mm, and 7.7\,Mm, respectively.
Inverted-Y shaped structures and moving blobs can be identified in some events. 
A subset of these events also reveal signatures in the HRI$_{Ly\alpha}$ (\hi\ \lya\ at 1216 \AA) images and the extreme ultraviolet images taken by the Atmospheric Imaging Assembly onboard the Solar Dynamics Observatory.
Our differential emission measure analysis suggests a multi-thermal nature and an average density of $\sim$1.4$\times$10$^{9}$ cm$^{-3}$ for these microjets.
Their thermal and kinetic energies were estimated to be $\sim$3.9$\times$10$^{24}$\,erg and $\sim$2.9$\times$10$^{23}$\,erg, respectively,
  which are of the same order of the released energy predicted by the nanoflare theory.
Most events appear to be located at the edges of network lanes and magnetic flux concentrations, suggesting that 
these coronal microjets are likely generated by magnetic reconnection between small-scale magnetic loops and the adjacent network field.

\end{abstract}

\section{Introduction}
\label{sec:intro}

Solar jets are one type of pervasive and explosive phenomena on the Sun, 
  usually appearing as a footpoint brightening followed by a nearly collimated plasma ejection \citep[e.g.,][]{2016SSRv..201....1R,2021RSPSA.47700217S,2018ApJ...861..108Z}.
Coronal jets are often observed in extreme ultraviolet \citep[EUV, e.g., ][]{2009SoPh..259...87N,2011RAA....11.1229Y,2012SoPh..281..729S,2014A&A...562A..98C,2018ApJ...854..155K,2019ApJ...885L..15K,2019ApJ...873...93K} and X-ray passbands \citep[e.g.,][]{1996PASJ...48..123S,2019ApJ...871..220S} in active regions (ARs) and coronal holes.
These coronal jets normally appear as anemone jets \citep[i.e., inverted-Y or $\lambda$ shape, e.g.,][]{2007Sci...318.1580C,2009SoPh..259...87N} or 
  two-sided-loop jets \citep[e.g.,][]{2017ApJ...845...94T,2020MNRAS.498L.104W}.
The anemone jets can be divided into blowout jets and standard jets \citep[e.g.,][]{2010ApJ...720..757M}.
The obvious differences are that a blowout jet often hosts two bright points (only one bright point at a standard one's base) and often involves a twisting mini-filament at its base \citep{2010ApJ...720..757M,2018ApJ...859....3M,2015Natur.523..437S,2017ApJ...844L..20Z,2019ApJ...887..239Y}.
Observational studies show that the coronal jets are often associated with or occur above satellite sunspots \citep{2008A&A...478..907C,2015ApJ...815...71C,2017PASJ...69...80S,2018ApJ...861..105S,2020ApJ...891..149P}, 
  coronal bright points \citep[CBPs,][]{2007PASJ...59S.751C,2016Ap&SS.361..301L,2019LRSP...16....2M}, 
  mini-filaments \citep{2016ApJ...821..100S,2016ApJ...830...60H,2018ApJ...854..155K,2019ApJ...873...93K}, or sigmoidal structures \citep{2016ApJ...817..126L,2014A&A...561A.104C}.
Coronal jets have been frequently found to be closely related to flux emergence, flux convergence,  and/or cancellation, indicating their generation by magnetic reconnection \citep{2007A&A...469..331J,2010A&A...519A..49H,2012ApJ...745..164S,2016ApJ...832L...7P}.
The brightening of the loops at the bases of some jets is found to be associated with a pattern of Doppler blue-to-red shifts, indicating an outbreak of siphon flow along the newly reconnected and submerging closed loop during the interchange magnetic reconnection \citep{2010A&A...519A..49H}.
The lengths, widths, projected speeds, lifetimes, temperatures, and densities of coronal jets are mostly 10$-$400\,Mm, 5$-$100\,Mm,
  10$-$1000\,\kms, tens of minutes, 3$-$8 MK, and 0.7$-$4.0$\times$10$^{9}$ cm$^{-3}$, respectively \citep{2000ApJ...542.1100S,2007PASJ...59S.771S,2016A&A...589A..79M,2018ApJ...868L..27P,2015A&A...579A..96P}.

Jet-like features have also been observed in the chromosphere and transition region (TR). These low-temperature jets are usually smaller than the coronal jets, and they appear as anemone jets in plage regions \citep[e.g.,][]{2007Sci...318.1591S,2021ApJ...913...59W}, penumbral jets \citep[e.g.,][]{2007Sci...318.1594K,2016ApJ...816...92T,2020A&A...642A..44H},
  light bridge jets \citep[e.g.,][]{2018ApJ...854...92T,2017AA...597A.127B,2017ApJ...848L...9H,2019ApJ...870...90B}, flare ribbon jets \citep{2019PASJ...71...14L}, cool polar jet \citep{2011A&A...534A..62S}, mini- or nano-jets from rotating prominence structures \citep{2017ApJ...841L..13C,2021NatAs...5...54A},
 chromospheric spicules \citep[e.g.,][]{2007PASJ...59S.655D,2015ApJ...799L...3R,2019Sci...366..890S}, TR network jets \citep[e.g.,][]{2014Sci...346A.315T,2016SoPh..291.1129N,2019ApJ...873...79C,2019SoPh..294...92Q}, and TR explosive events \citep[TREEs, interpreted as bi-directional jets,][]{1997Natur.386..811I}.
Though smaller in size, many chromospheric/TR jets have morphologies that are similar to those of coronal jets. For instance, many of these small jets also reveal inverted-Y shaped structures, indicating that magnetic reconnection between small-scale magnetic loops and the background field might be a common formation mechanism for these jets \citep[e.g.,][]{2020RSPSA.47690867N}.
The lengths, widths, velocities, and lifetimes of these chromospheric/TR jets have been found to be 1$-$11\,Mm, 100$-$400\,km, 5$-$250\,\kms, and 20$-$500\,s, respectively.
Due to their high occurrence rate and ubiquity, these small-scale jets have been suggested to contribute significantly to the heating of the upper solar atmosphere and origin of the solar wind \citep[e.g.,][]{2011Sci...331...55D,2014Sci...346A.315T,2019Sci...366..890S}.

With recent high-resolution observations taken by the Extreme Ultraviolet Imager \citep[EUI,][]{2020A&A...642A...8R}
  onboard Solar Orbiter \citep[SO,][]{2020A&A...642A...1M}, 
  we report the smallest coronal jets ever observed in the quiet Sun and investigate their physical properties.
We describe our observations in Section\,\ref{sec:obs}, present the analysis results in Section\,\ref{sec:res}, 
  discuss the results in Section\,\ref{sec:dis} and summarize our findings in Section\,\ref{sec:sum}.

\begin{table*}[htp]
\centering
\renewcommand\tabcolsep{3.pt}
\caption{Detailed information of the EUI datasets}
\begin{tabular}{c c c c c c c c }
\hline
Dataset&Date           &Time                &Distance from&Pixel size &Time &\lya ~data & SDO data \\
             &                    & [UT]               &Sun [AU]      &  HRI$_{EUV}$ (HRI$_{Ly\alpha}$) [Mm]  &  cadence [s]                 &   available?     &   available?    \\\hline
1            & 2020-05-20 & 21:12$-$22:09 & 0.61       & 0.22 (/)       & 10                 & /     & yes     \\
2            & 2020-05-21 & 16:04$-$17:30 & 0.60       & 0.22 (/)       & 10 (60)                & /     & yes     \\
3            & 2020-05-30 & 14:46$-$14:50 & 0.56       & 0.19 (0.21)        & 5                   & yes  &yes      \\
4            & 2020-10-19 & 19:44$-$20:31 & 0.99       & 0.35 (/)       & 12                 & /     &/         \\
5            & 2020-10-22 & 14:23$-$14:36 & 0.98       & 0.35 (/)       & 10                 & /     &/         \\
6            & 2020-11-19 & 11:52$-$14:31 & 0.92       & 0.32 (0.68)      & 15                 & yes  &/         \\\hline
\end{tabular}
\\
\tablenotetext{}{\textbf{For dataset\,2}, the regular time cadence is 10 s covering most micojets, whereas the time cadence of 60 s covers two microjets (Jet-09 and Jet-12). \textbf{For dataset\,6}, the HRI$_{Ly\alpha}$ images were binned with 2$\times$2 pixels.}
\label{datasets}
\end{table*}

\section{Observations}
\label{sec:obs}

The primary data\footnote{https://doi.org/10.24414/z2hf-b008} we used were taken by the two High Resolution Imagers (HRI) of EUI, 
  the passbands of which are centered at 174 \AA\ (HRI$_{EUV}$: dominated by Fe~{\sc{x}} and Fe~{\sc{ix}} lines) and 
  1216 \AA\ (HRI$_{Ly\alpha}$: dominated by the \lya\ line of Hydrogen), respectively \citep{2021arXiv210403382B}. 
The HRI$_{EUV}$ passband has a peak response at the temperature of  $\sim$1 MK \citep{2021arXiv210410940C}, and thus samples plasma in the corona. We mainly used HRI$_{EUV}$images to search for coronal microjets.
Six datasets taken on 2020 May 20, May 21, May 30, Oct 19, Oct 22, and Nov 19 were analyzed. 
The detailed information of these datasets is shown in Table\,\ref{datasets}. 

 We also analyzed simultaneous observations taken by the Atmospheric Imaging Assembly \citep[AIA,][]{2012SoPh..275...17L} and 
  the Helioseismic and Magnetic Imager \citep[HMI,][]{2012SoPh..275..207S}
  onboard the Solar Dynamics Observatory \citep[SDO,][]{2012SoPh..275....3P}
  on 2020 May 20, May 21, and May 30, when the field of views (FOVs) of EUI HRIs were also seen by SDO. 
We used the AIA 1600 \AA\ images at a 24 s cadence and 
  the 304 \AA, 171 \AA, 193 \AA, 211 \AA, 131 \AA, 335 \AA, and 94 \AA\ images at a 12 s cadence 
  to investigate the thermal property of the coronal microjets.
These AIA images have a pixel size of 430 km and a resolution of about 1 Mm. 
To investigate their origin, we also examined the HMI line-of-sight (LOS) magnetic field data at a 45 s cadence.
The pixel size of the magnetograms is 360 km.

Since the HRI$_{EUV}$ and the AIA 171 \AA\ images have similar temperature responses, we first aligned them through a linear Pearson correlation. And other AIA and HMI images were then aligned with the AIA 171 \AA\ images using the aia\_prep.pro routine available in SolarSoftware (SSW).
For aligning the HRI$_{EUV}$ and HRI$_{Ly\alpha}$ images, several CBPs in the images were used as the referent features.
Due to the difference in the heliocentric distance of SO and SDO, 
  the solar radiation emitted at a particular time reaches the two spacecraft at different times.
To avoid confusion in further analysis, we converted the observation times of all telescopes into the light emitting times on the Sun, which are listed in Table\,\ref{datasets}.

\section{Results}
\label{sec:res}

\begin{figure*}
\centering
\includegraphics[trim=0.0cm 1.0cm 0.0cm 0.0cm,width=1.0\textwidth]{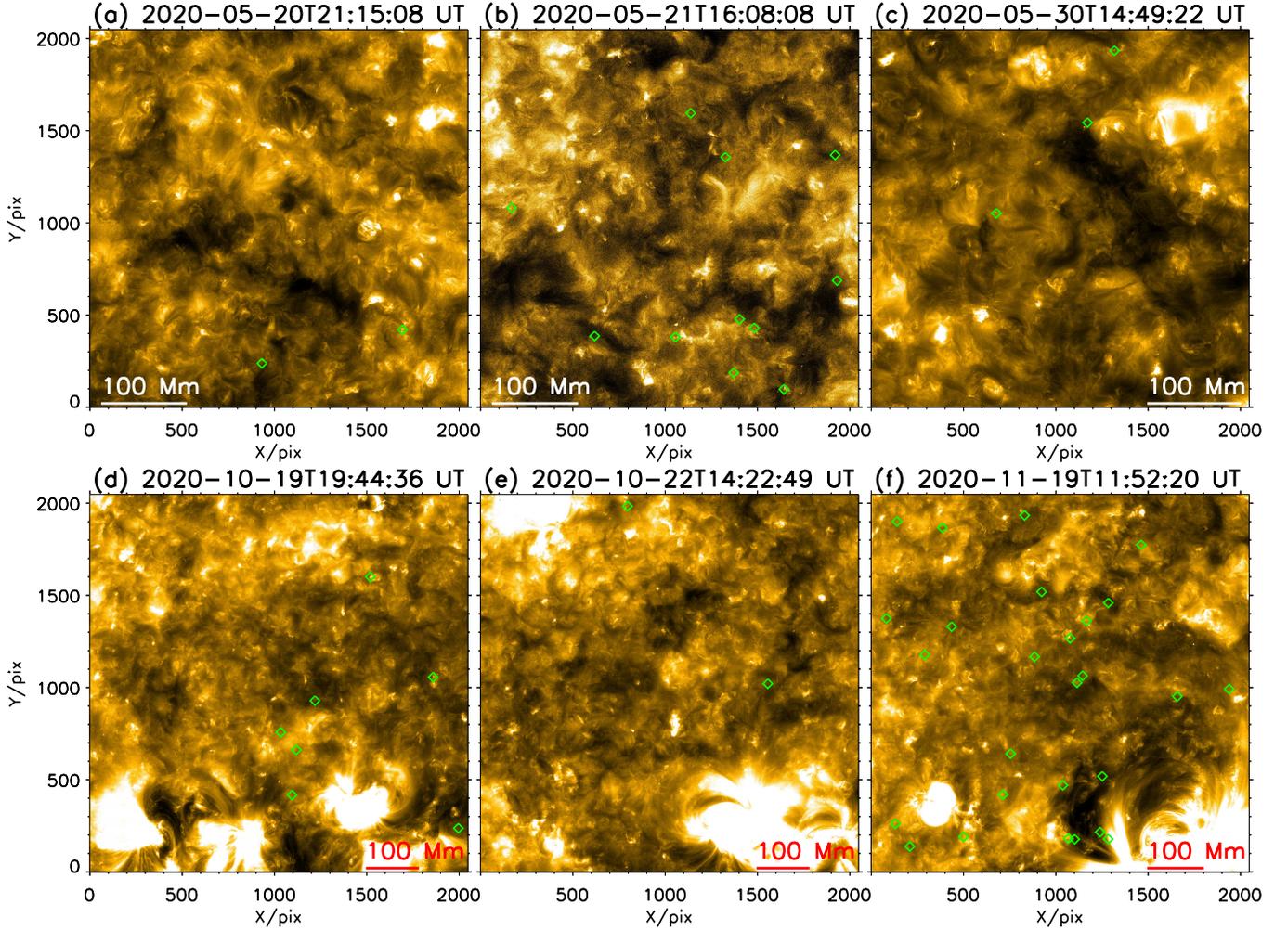}
\caption{ Overviews for all the six datasets in the HRI$_{EUV}$ passband.
The green open diamonds in each panel mark the footpoint locations of the identified coronal microjets.
}
\label{fig:overview}
\end{figure*}

Figure\,\ref{fig:overview} shows an overview for each of the six datasets in the HRI$_{EUV}$ passband.
The Sun was very quiet on these days, and the FOVs mainly include wide-spreading CBPs and diffuse regions in between. Several ARs appear near the edges of the FOVs in datasets 4-6; these ARs were saturated due to their high radiance and were excluded when searching for coronal jets.
The high spatial and temporal resolutions allow us to investigate various types of small-scale transient activity. Previous investigations have identified numerous small brightenings termed campfires from these HRI$_{EUV}$ images \citep{2021arXiv210403382B}. Here we focus on small-scale collimated jet-like features, which we call coronal microjets. 

In total, we identified 52 coronal microjets from the six HRI$_{EUV}$ image sequences. The footpoint locations of all these jets are marked in Fig.\,\ref{fig:overview}.
The detailed information such as he staring times of their occurrence on the Sun, durations, and pixel locations are listed in Table\,\ref{paras}.
There are 2 events in each of the datasets\,1,\,3 and 5,  11 events in dataset\,2, 7 events in dataset\,4, and 27 events in dataset\,6. 
Figure\,\ref{fig:overview} shows that most of these microjets are located in the quiet-Sun regions that are far away from the ARs.
A few of them are located close to but not inside the ARs.

\subsection{Morphology and parameters of the coronal microjets in HRI$_{EUV}$}
\label{subsec:jets_in_174}

\begin{figure*}
\centering
\includegraphics[trim=0.0cm 0.8cm 0.0cm 0.0cm,width=1.0\textwidth]{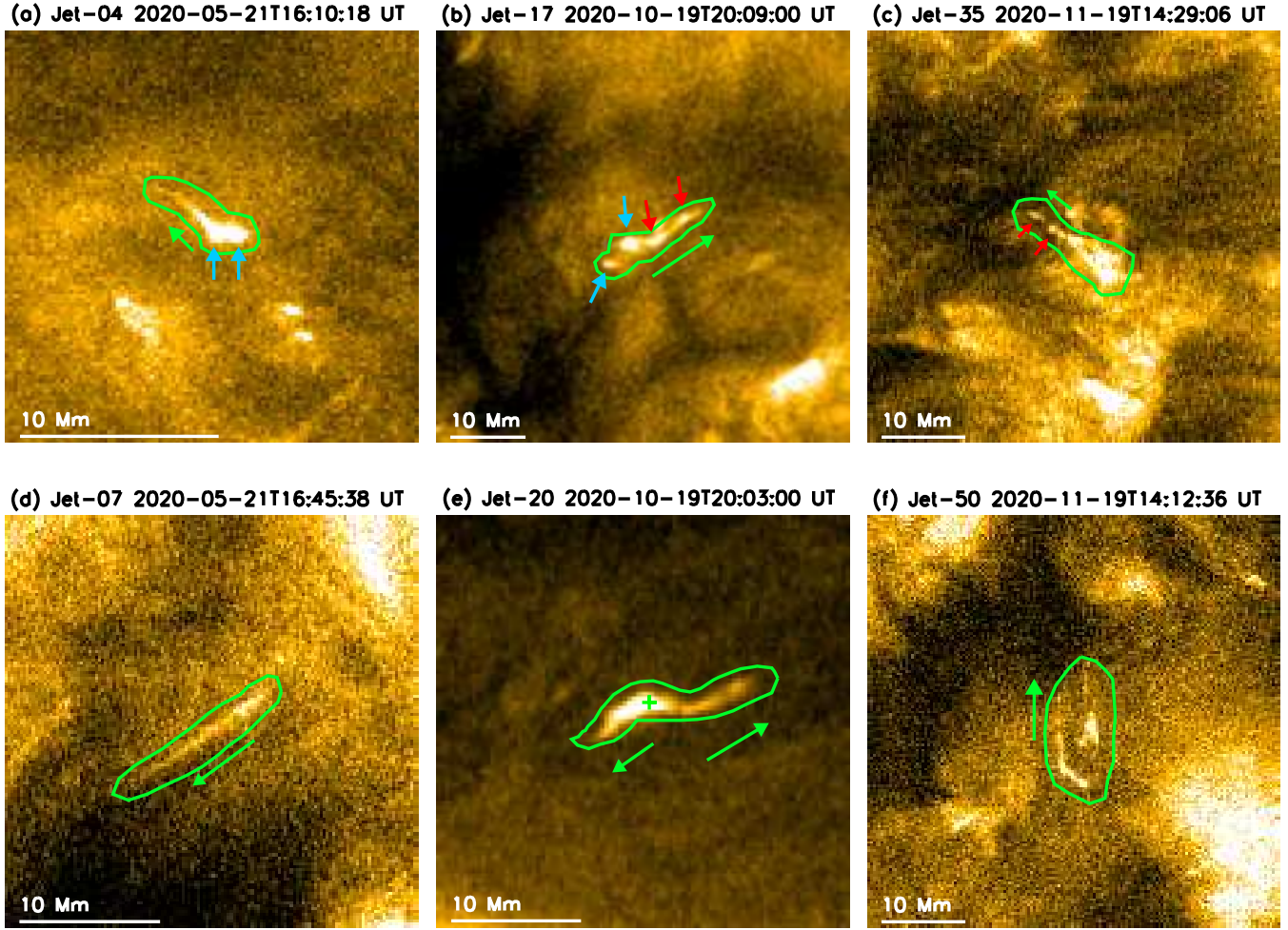}
\caption{Morphologies of Jet-04, Jet-17, Jet-35, Jet-07, Jet-20, and Jet-50 in the HRI$_{EUV}$ images.
The green contours outline the microjets, and the green arrows indicate the propagation directions of the microjets.
In (a) and (b), the blue arrows mark the footpoints of the microjets.
In (b) and (c), the red arrows indicate some blobs within the microjets.
In (e), the green plus marks the location of the starting brightening in Jet-20.
An animation of this figure is available, showing the evolution of these six microjets in the HRI$_{EUV}$ images.
It includes six panels covering durations of 2 min from 16:09:38 UT to 16:11:38 UT on 2020 May 21, 10.6 min from 20:02:24 UT to 20:13:00 UT on 2020 October 19, 8.3 min from 14:23:51 UT to 14:32:06 UT on 2020 November 19, 5.7 min from 16:41:28 UT to 16:47:08 UT on 2020 May 21, 3.3 min from 20:00:41 UT to 20:04:00 UT on 2020 October 19, and 7.8 min from 14:07:51 UT to 14:15:36 UT on 2020 November 19, respectively.
In this animation, the green arrows mark the microjets, the red arrows in (b) and (c) mark the blobs, and the white arrows in (f) denote two dark regions that appear after the disappearance of Jet-50.
}
\label{fig:example}
\end{figure*}

The identified microjets are generally preceded by clear brightenings at their footpoints in the HRI$_{EUV}$ images. They mostly appear as collimated structures shooting up from the footpoint brightenings. Their morphology is very similar to the known larger-scale coronal jets.
As an example, we show 6 coronal microjets in Figure\,\ref{fig:example}.

Figure\,\ref{fig:example}(a) gives a closer look at Jet-04. To highlight the jet, we manually drew a green contour enclosing the jet structure. 
The green arrow indicates the propagation direction of the jet.
This jet appears to consist of two footpoints and a collimated jet body, forming an inverted-Y shaped structure.
Figure\,\ref{fig:example}(b) gives a closer look at Jet-17.
In addition to the two footpoints at the base and the quasi-collimated jet body, two blobs marked by the red arrows in Figure\,\ref{fig:example}(b) can be clearly seen in the jet. These moving blobs appear to be similar to those reported in large-scale coronal jets \citep{2014A&A...567A..11Z,2016SoPh..291..859Z,2018ApJ...854..155K,2019ApJ...885L..15K,2019ApJ...873...93K}.
Two blobs can also be identified from Jet-35, as marked by the red arrows in Figure\,\ref{fig:example}(c). For a detailed evolution of these moving blobs, we refer to the associated animation of Figure\,\ref{fig:example}. The jet shown in Figure\,\ref{fig:example}(d) (Jet-07) is a highly collimated one. It includes a brighter, confined base and a fuzzy but identifiable jet body.
Figure\,\ref{fig:example}(e) shows a bi-directional jet (Jet-20). We see that plasma is ejected towards the opposite directions from the jet base, which is marked by the plus symbol.
The last example is Jet-50, shown in Figure\,\ref{fig:example}(f).
Two microjets occur close to each other simultaneously. The footpoints of these two jets appear to be kinked or reveal two branches, possibly indicating an inverted-Y shape morphology.
From the associated animation, we can see the appearance of two dark regions following the disappearance of Jet-50, which might be similar to the coronal dimmings after eruptions of coronal mass ejections \citep[e.g.,][]{1997ApJ...491L..55S,2020Innov...100059X}.
Moreover, from Figure\,\ref{fig:example} and the associated animation we see that most microjets appear as isolated structures without any indication of pre-erupion mini-filaments at their bases, which is different from many previously reported coronal jets in ARs and coronal holes.

\begin{figure*}
\centering
\includegraphics[trim=0.0cm 0.6cm 0.0cm 0.0cm,width=1.0\textwidth]{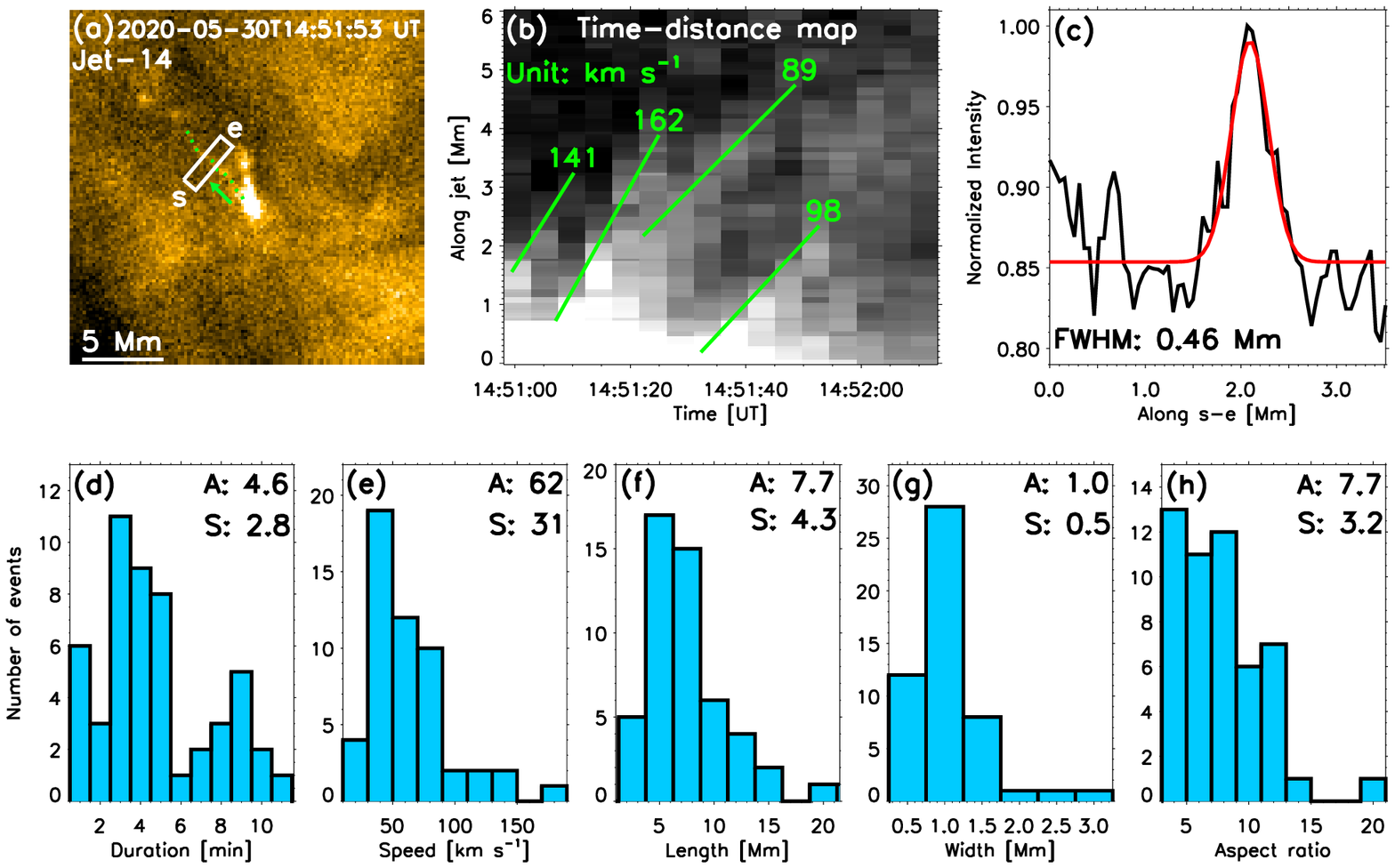}
\caption{Parameters of the coronal microjets.
(a) An HRI$_{EUV}$ image showing Jet-14 at 14:51:35 UT on 2020 May 30.
The green arrow indicates the propagation direction, and the green dotted line marks the trajectory of the ejected plasma.
(b) The time-distance map for the green dotted line in (a). Four tracks of ejected plasma and the corresponding projected speeds in the unit of \kms\ are marked by the green lines and the numbers, respectively.
(c) The intensity variation along the long side of the rectangle s$-$e shown in (a).
The original intensity profile and the Gaussian fit are indicated by the black and red curves, respectively.
(d)$-$(h) Distributions of the parameters for the microjets.
In each panel, `A' and `S' represent the average value and standard deviation, respectively.}
\label{fig:paras}
\end{figure*}

We measured the physical parameters for these coronal microjets, i.e. the duration, projected speed, length, width, and aspect ratio, which are listed in Table\,\ref{paras}. 
The lifetime of each event was calculated as the time difference between the first appearance of the footpoint brightening and the disappearance of the jet body.
For several events either the first appearance of the footpoint brightening or the disappearance of the jet body was not captured by EUI, we thus 
cannot accurately estimate their full lifetimes. Instead, we registered the existing periods of these jets in the HRI$_{EUV}$ images (see Table\,\ref{paras}).
We marked the trajectory of each microjet and calculated the speed from the corresponding time-distance map.
As an example, Figure\,\ref{fig:paras}(a) and (b) show the trajectory and the time-distance map for Jet-14, respectively.
Four slant stripes, marking four episodes of plasma ejection in Jet-14, can be seen from Figure\,\ref{fig:paras}(b). The speeds for the four ejections were estimated from the slopes, which range from 89 to 162 \kms. We took the average speed 123 \kms\ as the speed of this jet for further analysis. We would like to mention that most microjets only show one episode of ejection, and in such cases there is no need to average the speeds. 
We calculated the length of a microjet when it was fully developed and clearly resolved.
To estimate the width of a microjet, we put a narrow white rectangle perpendicular to the microjet, as exemplified in Figure\,\ref{fig:paras}(a). We then plotted
the intensity along the long side of the rectangle and obtained the intensity profile across the jet, e.g., the black line in Figure\,\ref{fig:paras}(c).
We applied a Gaussian fit to the intensity profile, and the full-width-at-half-maximum (FWHM) was taken as the width of the microjet.
Finally, the aspect ratio of the microjet could be calculated as the ratio of the length and width.

Figure\,\ref{fig:paras}(d)$-$(h) show the distributions of their parameters.
Their durations are less than 11 min, with an average of $\sim$4.6 min.
Their projected speeds are mostly less than 100 \kms\, with an average value of $\sim$62\,\kms.
These microjets have lengths ranging from 2 to 20 Mm, with an average of $\sim$7.7 Mm. Their widths, obtained as the FWHM of a Gaussian function fitted to the intensity profile, are in the range of 0.3 to 3.0 Mm, and the average value is $\sim$1.0 Mm.
The aspect ratios of the microjets range from 3 to 20, and the average is 7.7.

\subsection{Response in the HRI$_{Ly\alpha}$ images}
\label{subsec:lya}

\begin{figure*}
\centering
\includegraphics[trim=0.0cm 0.5cm 0.0cm 0.0cm,width=0.78\textwidth]{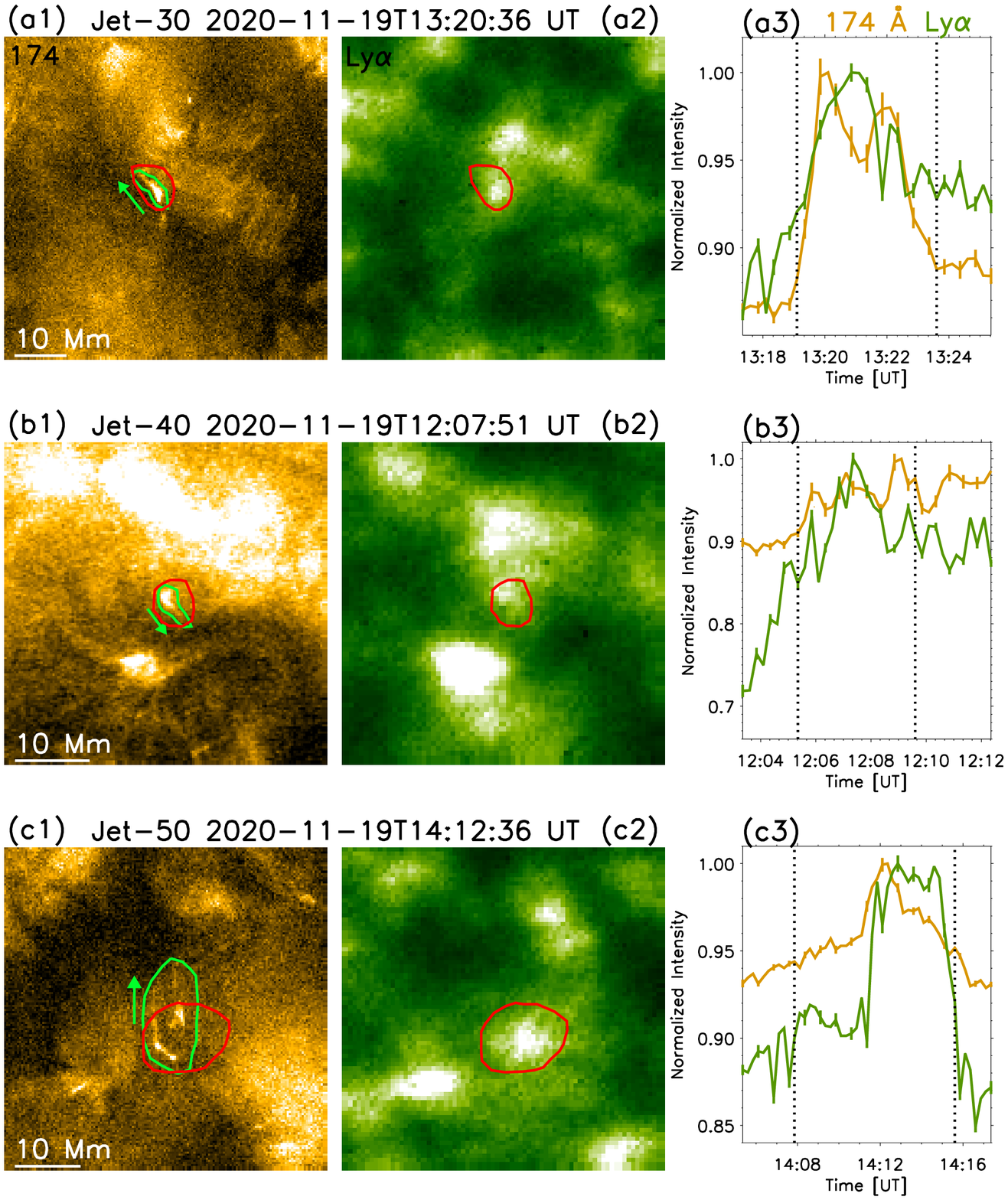}
\caption{Responses of Jet-30, Jet-40, and Jet-50 in the HRI$_{Ly\alpha}$ images.
Left and middle columns: snapshots of three microjets in the HRI$_{EUV}$ and HRI$_{Ly\alpha}$ images.
The green contours outline the microjets, and the green arrows represent the propagation directions of the jets.
The red contours were used to obtain the average intensity variations at 174 \AA\ and \hi\ \lya, 
  and the corresponding light curves are shown in the right column.
In the right column, the vertical bars indicate the standard errors of the mean values, and the vertical dashed lines represent the starting and ending times of the events.
}
\label{fig:lya}
\end{figure*}

We also examined responses of the identified coronal microjets in the HRI$_{Ly\alpha}$ images.
Datasets 3 (3 events) and \,6 (27 events) have simultaneous observations in the HRI$_{Ly\alpha}$ passband.
We found that 11 out of the 30 jets (see the details in Table\,\ref{paras}) have emission responses at the jet locations in the HRI$_{Ly\alpha}$ images, and three examples are shown in Figure\,\ref{fig:lya}.
We chose one region around the footpoint of each microjet
  (red contours in the left and middle columns of Figure\,\ref{fig:lya}) and 
  plotted the light curves of the total intensities at 174 \AA\ and \hi\ \lya\ in the right column of Figure\,\ref{fig:lya}.
The two light curves reveal a similar trend for these three microjets, indicating that emissions in the two passbands are mainly contributed by the same process. These microjets with clear response in \lya\ might be generated at lower heights as compared to others.  

By comparing the left and middle columns of Figure\,\ref{fig:lya},
  we can see that these microjets are located at or near the network lanes that are characterized by the enhanced emission in the HRI$_{Ly\alpha}$ images.
Actually, at least half of the 30 microjets (including the three microjets shown in Figure\,\ref{fig:lya}) appear to be located at the edges of the network lanes. Only one microjet (Jet-43, see Table\,\ref{paras}) is not near the network lane.

\begin{figure*}
\centering
\includegraphics[trim=0.0cm 0.8cm 0.0cm 0.0cm,width=0.9\textwidth]{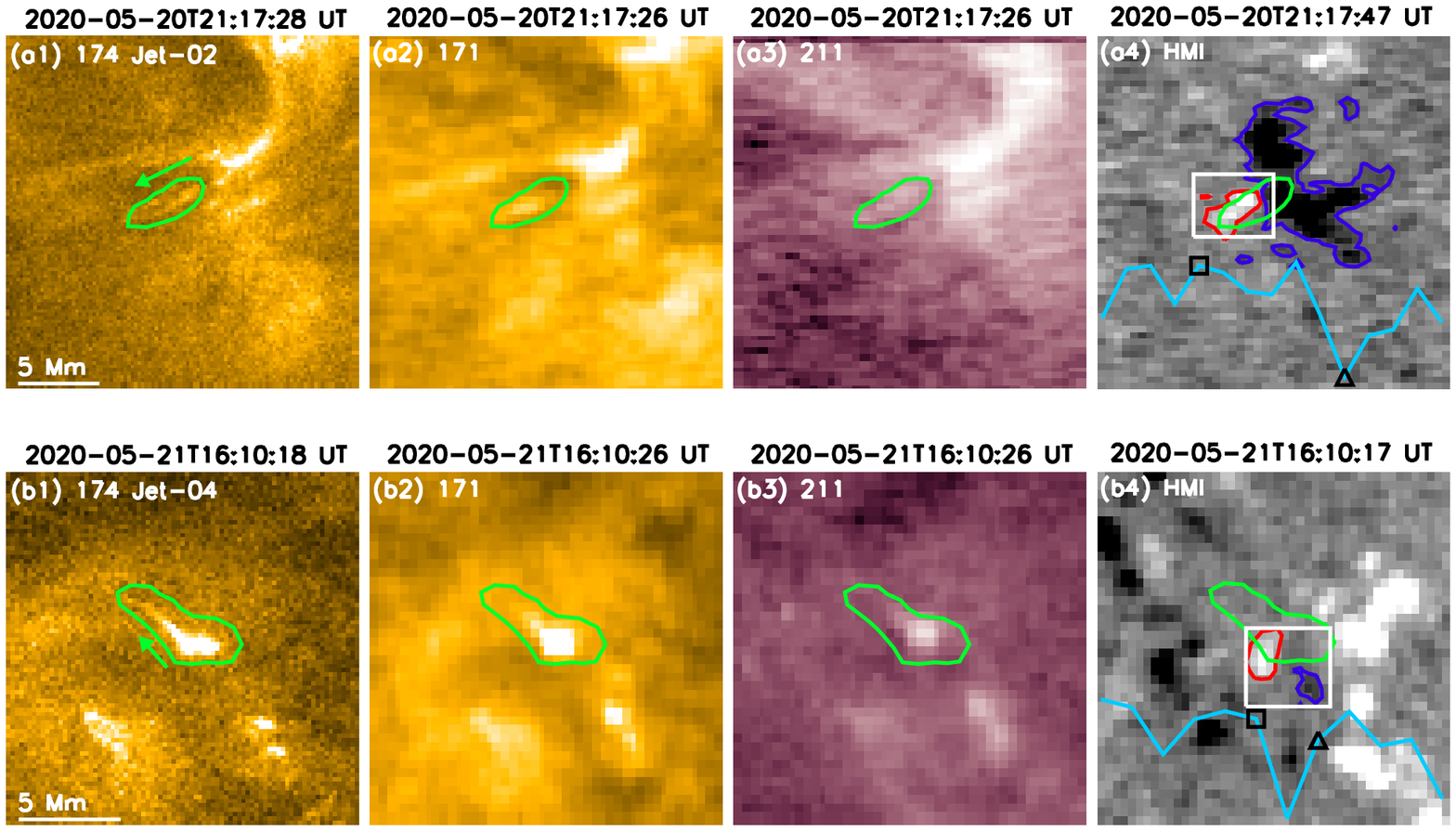}
\caption{From left to right: Images of the HRI$_{EUV}$, AIA 171\,\AA, 211 \AA, and HMI LOS magnetic field for Jet-02 (upper panels) and Jet-04 (lower panels).
The green arrows indicate the propagation directions of the jets.
The green contours mark the locations of the microjets.
The red and blue contours in (a4) and (b4) represent magnetic field strength with the levels of $\pm$10 G, respectively.
The HMI magnetograms are saturated at $\pm$50\,G.
In (a4) and (b4), the overplotted cyan curves show the normalized variations of the minority-polarity (positive for (a4) and negative for (b4)) magnetic fluxes in the white boxes.
The black squares and triangles represent the starting and ending times for the events, respectively.
}
\label{fig:sdo}
\end{figure*}

\subsection{Response in the AIA images and the associated magnetic field}
\label{subsec:mgrec}

The FOVs of datasets \,1 (2 events), \,2 (11 events), and \,3 (3 events)
  have also been observed by AIA and HMI.
We thus examined the responses of these events in the AIA UV and EUV images, and investigated the possible origin of these jets using the HMI line-of-sight magnetograms.
We did not find any signature of these events in the AIA 1600 \AA\ images.
Many microjets reveal obvious signatures in the AIA 304 \AA, 171 \AA, 193 \AA, and 211 \AA\ passbands,
  and for some cases weak signatures can be identified from the AIA 335 \AA\ and 131 \AA\ passbands (Table\,\ref{paras}).
As an example, Figure\,\ref{fig:sdo} shows two events, Jet-02 and Jet-04, 
  in the HRI$_{EUV}$, AIA 171 \AA, and 211 \AA\ images.
These two events reveal clear signatures in the AIA 171 \AA\ and 211 \AA\ images, especially around their footpoint regions.
The jet signatures are fuzzier in the AIA EUV passbands 
   than in the HRI$_{EUV}$ images.
Without the high-resolution HRI$_{EUV}$ observations, it would be difficult to identify these coronal microjets from the AIA EUV passbands.

With the HMI magnetograms, we have examined the magnetic field structures around the coronal microjets.
We found that ten events are located in regions with opposite magnetic polarities (see the details in Table\,\ref{paras}).
In Figure\,\ref{fig:sdo}\,(a4) and (b4), we show the line-of-sight magnetic field around Jet-02 and Jet-04.
Jet-02 is obviously located above a region with opposite polarities, where the negative polarity is dominated.
The footpoint region of Jet-04 is also a mixed-polarity region, where a weak negative flux is close to the strong positive flux of the network.
In some cases, there are more than one patches of the minor polarity around their footpoints.
These events were also considered to be related to opposite-polarity fluxes if one patch consists of more than 4 pixels 
  with a field strength larger than 10\,G.

The resolution of the HMI magnetograms is not high enough to reveal clear flux changes of the magnetic field associated with most of the microjets. However, we did find clear magnetic flux changes for a few microjets. We present the variations of the minority-polarity magnetic fluxes for two events in Figure\,\ref{fig:sdo}\,(a4) and (b4), where we can see clear signatures of flux decrease between the starting and ending times of each event. In addition, the positive magnetic flux (not shown in Figure\,\ref{fig:sdo}\,(b4)) associated with Jet-04 reveals an obvious decrease during the occurrence of the jet, similar to the negative flux.
Magnetic cancellation is also associated with some coronal jets \citep[e.g.,][]{2015ApJ...815...71C,2017ApJ...840...54C}.
These observational signatures indicate that magnetic cancellation is likely associated with at least some microjets.

Using the coaligned AIA 1600 \AA\ images and HMI magnetograms, we can also find that almost all of the 16 microjets identified from datasets 1, 2 and 3 originate from the bright network lanes or the relatively strong magnetic flux concentrations.
The footpoints of some microjets are clearly located at the edges of magnetic flux concentrations.
As mentioned above, a similar result was also found from the HRI$_{Ly\alpha}$ images (see the details in Table\,\ref{paras}).

\subsection{DEM analysis}
\label{subsec:details}

\begin{figure*}
\centering
\includegraphics[trim=0.0cm 0.3cm 0.0cm 0.0cm,width=0.99\textwidth]{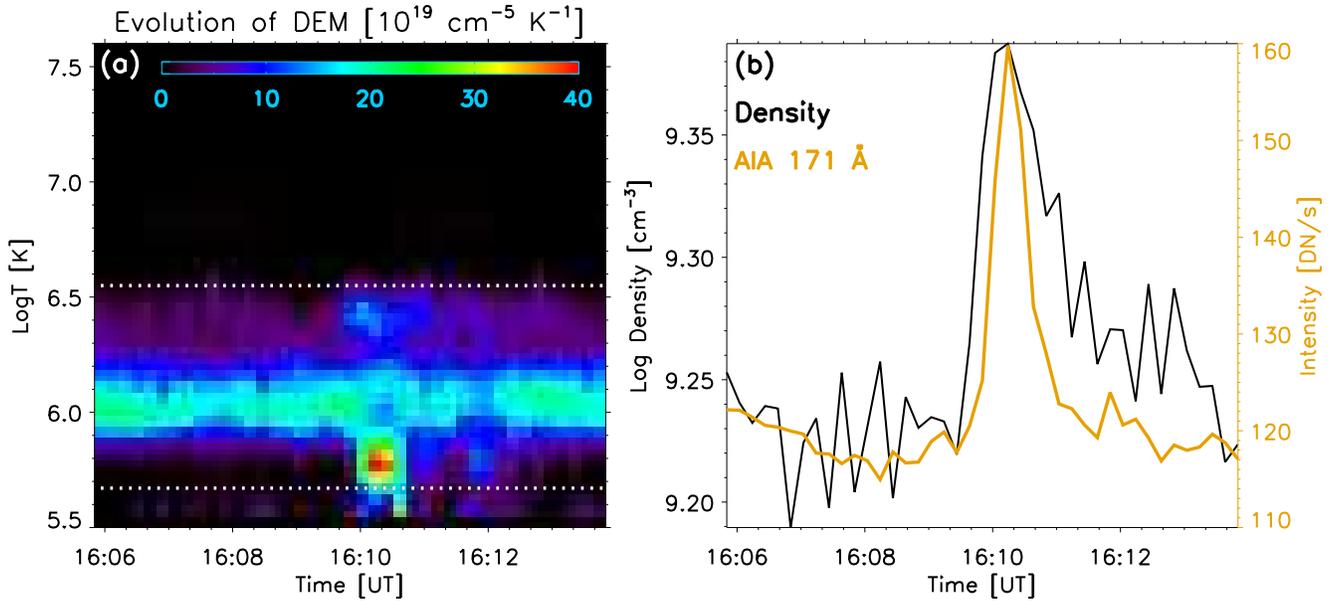}
\caption{DEM analysis for Jet-04.
(a) Evolution of the DEM. The two white dotted lines indicate the temperature range within which the DEM is obviously enhanced when the jet occurs.
(b) Temporal evolution of the electron density (black) and AIA 171 \AA\ intensity (yellow) averaged over the green contour in Figure\,\ref{fig:sdo}\,(b2).}
\label{fig:dem}
\end{figure*}

For the coronal microjets observed in 2020 May (except Jet-15 without emission response in the AIA EUV passbands),
  we performed a differential emission measure (DEM)
  analysis \citep{2015ApJ...807..143C,2018ApJ...856L..17S,2020ApJ...898...88X,2021Innov...200083S} to investigate their plasma properties with observations 
  in the AIA 94 \AA, 131 \AA, 171 \AA, 193 \AA, 211 \AA, and 335 \AA\ passbands.
As an example, we averaged the intensities within the green contour in Figure\,\ref{fig:sdo}\,(b2) and then performed the DEM analysis for Jet-04. We show the temporal evolution of the resultant DEM in Figure\,\ref{fig:dem}\,(a).
From 16:09:40 UT to 16:10:20 UT, Jet-04 appears in the AIA EUV passbands and 
  the DEM in the jet region is obviously enhanced in a broad temperature range of log (T/K)\,=\,5.65$-$6.55.
We also calculated its EM-weighted temperature and obtained a value of log (T/K)\,=\,$\sim$6.20, which is similar to that reported by \citet{2015A&A...579A..96P} using the filter-ratio technique.
A similar behavior was also found for other coronal microjets.
This indicates the multi-thermal nature for the plasma in these microjets.

Taking the width (0.4\,Mm) of Jet-04 as the LOS integration depth, we can estimate the electron density of Jet-04
  by integrating the DEM over the temperature range of log (T/K)\,=\,5.65$-$6.55.
Figure\,\ref{fig:dem}\,(b) shows the temporal evolution of the density, which closely matches the light curve of the AIA 171 \AA\ intensity.
From Figure\,\ref{fig:dem}\,(b), we can see that the density of Jet-04 is $\sim$2.2$\times$10$^{9}$ cm$^{-3}$.
The densities for other microjets are shown in Table\,\ref{paras}.
The average density for these quiet-Sun microjets is $\sim$1.4$\times$10$^{9}$ cm$^{-3}$, which is roughly one order of magnitude lower than the typical density of coronal jets observed in ARs \citep[e.g.,][]{2008A&A...481L..57C,2012ApJ...748..106T}. 

Since we have obtained the physical parameters (spatial scale, projected speed, temperature, density),  we can calculate the thermal energy 
  (E$_{t}$ = 3N$_{e}$k$_{B}$TV) and kinetic energy (E$_{k}$ = 0.5N$_{e}$m$_{p}$$v^{2}$V) for the coronal microjets \citep[e.g.,][]{2014ApJ...790L..29T,2020ApJ...899...19C}. Here N$_{e}$, m$_{p}$, k$_{B}$, $v$, V, and T are the electron number density, proton mass, Boltzmann constant, velocity, volume, and temperature of a microjet. 
By assuming a coronal microjet with a cylindrical shape (a height of 7.7\,Mm and a cross section diameter of 1.0\,Mm), a projected speed of 62\,\kms,
  a mean coronal temperature of 10$^{6}$\,K, and an electron density of $\sim$1.4$\times$10$^{9}$ cm$^{-3}$, 
its thermal and kinetic energies were estimated to be $\sim$3.9$\times$10$^{24}$\,erg and $\sim$2.9$\times$10$^{23}$\,erg, respectively. 
Thus, the total energy of a typical microjet appears to be of the same order of the released energy predicted by the nanoflare theory \citep[$\sim$10$^{24}$\,erg,][]{1988ApJ...330..474P}.

\section{Discussion}
\label{sec:dis}
At first sight, our microjets appear to be different from the campfires recently discovered from the HRI$_{EUV}$ images. Campfires are characterized by small-scale brightenings in quiet-Sun regions \citep{2021arXiv210403382B}. Most campfires have a loop-like morphology, with a length smaller than 4 Mm and a length to width ratio of 1--5. While the microjets detected here generally have a much larger aspect ratio, 7.7$\pm$3.2. The lengths of microjets, 7.7$\pm$4.3 Mm are also considerably larger than those of most campfires. In addition, the microjets are characterized as plasma moving upward from a footpoint brightening, which is not typical for campfires. Another difference is that some coronal microjets also reveal a response in the HRI$_{Ly\alpha}$ images, whereas campfires generally show no signature in \lya.
Nevertheless, we have examined the same dataset (dataset\,3) used by \citet{2021arXiv210403382B} and found that 
  the coronal microjets we identified from dataset\,3 have also been identified as campfires by the detection method of \citet{2021arXiv210403382B}.
In several cases, a coronal microjet was identified as sequential campfires or spatially adjacent campfires.
Thus, we conclude that the identified coronal microjets are a peculiar subset of coronal campfires.
Different from the suggestion that many campfires might correspond to apexes of low-lying small-scale network loops heated at the coronal heights by component reconnection between different loops \citep{2021arXiv210410940C}, the coronal microjets are more likely to be generated by magnetic reconnection between small-scale magnetic loops and the ambient network field at lower heights.

Another group of small-scale dynamic events resulting from magnetic reconnection are TREEs \citep[e.g., ][]{1991JGR....96.9399D,1997Natur.386..811I,2006SoPh..238..313C}, which are characterized by non-Gaussian line profiles at TR temperatures \citep[e.g.,][]{1983ApJ...272..329B,2003A&A...403..287P}.
Observations have shown that TREEs are usually located at the boundary of magnetic network \citep[e.g., ][]{2003A&A...403..731M,2004A&A...427.1065T}.
Furthermore, they are often associated with jet-like structures \citep[e.g., ][]{2014ApJ...797...88H,2017MNRAS.464.1753H,2019ApJ...873...79C}.
Here we found that the bases of almost all coronal microjets are located at the network lanes, with many of them at the edges of the lanes.
Future investigations are required to understand whether the coronal microjets and TREEs are related.
 
With AIA observations, \citet{2014ApJ...787..118R} reported a new type of coronal jets, called `jetlets', which are smaller than the typical coronal jets.
They are often located at the bases of coronal plumes and associated with opposite magnetic polarities \citep[e.g.,][]{2015ApJ...807...71P,2018ApJ...868L..27P}. Sometimes they can also be identified from the TR images taken by the Interface Region Imaging Spectrograph \citep[IRIS,][]{2014SoPh..289.2733D}.
These jetlets have an average length of 27\,Mm and an average width of 3\,Mm, about three times larger than the coronal microjets we report here.
More recently, using observation from the High-resolution Coronal Imager 2.1 \citep[Hi-C 2.1,][]{2019SoPh..294..174R},
  \citet{2019ApJ...887L...8P} found six smaller jetlet-like structures with an average length of 9\,Mm, at edges of network lanes close to an AR.
Some of these jetlets appear to be the larger population of TR network jets \citep{2014Sci...346A.315T}. Some other jetlets might be similar to our quiet-Sun microjets, although our microjets are generally smaller and not associated with coronal plumes. More investigations are required to understand the relationship between the coronal microjets and jetlets. With observations of Hi-C 2.1, \citet{2019ApJ...887...56T} recently discovered some fine-scale surge/jet-like features in the core of an AR.  Future observations should be performed to examine whether these features are the AR counterpart of our microjets. 

Huge efforts have been made to investigate the physical nature and formation mechanisms of solar jets. Magnetic reconnection is generally believed to be the primary cause of solar jets. Numerical simulations of reconnection driven jets have reproduced many observational features of solar jets. For example, \citet{1995Natur.375...42Y} and \citet{1996PASJ...48..353Y} performed MHD simulations and reproduced both the anemone jets (inverted-Y shaped jets) and two-sided-loop jets (bidirectional jets).
These features, as well as the brightenings caused by plasma heating at the jet footpoints, have also been detected in some of our identified coronal microjets.
In reconnection regions, plasma blobs (or plasmoids) can form and move within the jets \citep[e.g.,][]{1995Natur.375...42Y,1996PASJ...48..353Y}.
In a high plasma-$\beta$ environment, the Kelvin$-$Helmholtz instability may develop and lead to the formation of vortex-like blobs \citep{2017ApJ...841...27N,2018RAA....18...45Z}.
On the other hand, the plasmoid instability could occur in the low plasma-$\beta$ case, and multi-thermal plasmoid blobs may form and move upward in the jets \citep{2017ApJ...841...27N}.
Considering that the corona has a low-$\beta$, the moving blobs identified in some microjets are more likely to be plasmoids.
The fact that many microjets are associated with mixed-polarity magnetic fluxes and located at the edges of network lanes favors the scenario of jet production by magnetic reconnection between small-scale magnetic loops and the ambient network field in the quiet-Sun regions.
This scenario appears to be consistent with that of the simulations performed by \citet{2013ApJ...777...16Y} and \citet{2018ApJ...852...16Y}.
In these simulations, magnetic reconnection occurs between closed field and the locally open network field, leading to the generation of an inverted-Y shaped jet and moving blobs in the simulated jet.

Some of these microjets might also be part of a continuum of jets that emanate from embedded bipoles, i.e., the three-dimensional (3D) fan-spine topology, consisting of a minority polarity surrounded by majority-polarity fluxes.
This scenario has been extensively studied through observational analysis \citep[e.g.,][]{2018ApJ...854..155K,2019ApJ...885L..15K,2019ApJ...873...93K} and 3D simulations
\citep[e.g.,][]{2009ApJ...691...61P,2015A&A...573A.130P,2016A&A...596A..36P,2016ApJ...820...77W,2016ApJ...827....4W,2017Natur.544..452W,2017ApJ...834...62K}. Mini-filament eruptions are often involved in this process. 
For the microjets reported here, no obvious signatures of small-scale (or tiny-, micro-) filaments were observed by the HRI$_{EUV}$ images. However, this might be due to the fact that the associated filament structures are too small to be resolved by EUI.
 
Our observations of numerous microjets in quiet-Sun regions might be consistent with the results of \citet{2016SoPh..291.1357W}, who suggested interchange reconnection as the coronal source of superhalo electrons \citep{2012ApJ...753L..23W}. Specifically, they proposed that
  small-scale interchange reconnections in quiet-Sun regions produce an upward-traveling population of accelerated electrons,
  which could escape into the interplanetary space and form the superhalo electrons measured in the solar wind.

\section{Summary}
\label{sec:sum}

We have identified 52 coronal microjets from the HRI$_{EUV}$ images taken on six different days.
These coronal microjets appear as quasi-collimated plasma ejections from clear brightenings in the quiet-Sun regions.
Some of these microjets reveal an inverted-Y shape and include moving blobs, which are similar to many previously known jets.
The footpoints of most microjets are located at the edges of network lanes and associated with mixed-polarity magnetic fluxes.
These observational features suggest that the coronal microjets are the small-scale version of coronal jets, which are likely generated by magnetic reconnection between small-scale magnetic loops and the ambient network field in the quiet-Sun regions.

We have measured various physical parameters of the coronal microjets, and found that 
  (1) the durations are mostly a few minutes, 
  (2) the projected speeds are mostly 40--90 \kms,
  (3) the average lengths and width are $\sim$7.7 Mm and $\sim$1.0 Mm, respectively,
  (4) the aspect ratios of the jets are generally 4--13.
We have examined the response of the coronal microjets in the HRI$_{Ly\alpha}$ images and found that 
  11 out of 30 jets show a response, possibly indicating the generation of these 11 microjets at lower heights compared to others.
When the AIA data are available, we found that almost all of the microjets have signatures in the AIA EUV passbands, 
  especially in the 304 \AA, 171 \AA, 193 \AA, and 211 \AA\ passbands.
Through a DEM analysis, we found that these coronal microjets are multi-thermal, and their average density is $\sim$1.4$\times$10$^{9}$ cm$^{-3}$.
 The thermal and kinetic energies of these coronal microjets were estimated to be 
  $\sim$3.9$\times$10$^{24}$\,erg and $\sim$2.9$\times$10$^{23}$\,erg, respectively, which fall into the energy range of coronal nanoflares.

\acknowledgments
This work was supported by NSFC grants 11825301 and 11790304, 41874200, and the Strategic Priority Research Program of CAS (grant XDA17040507).
H.C.C. was supported by the National Postdoctoral Program for Innovative Talents (BX20200013) and China Postdoctoral Science Foundation (2020M680201).
Solar Orbiter is a space mission of international collaboration between ESA and NASA, operated by ESA. The EUI instrument was built by CSL, IAS, MPS, MSSL/UCL, PMOD/WRC, ROB, LCF/IO with funding from the Belgian Federal Science Policy Office (BELSPO/PRODEX PEA  4000112292); the Centre National d'Etudes Spatiales (CNES); the UK Space Agency (UKSA); the Bundesministerium für Wirtschaft und Energie (BMWi) through the Deutsches Zentrum für Luft- und Raumfahrt (DLR); and the Swiss Space Office (SSO).
AIA and HMI are instruments onboard the Solar Dynamics Observatory, a mission for NASA's Living With a Star program.

\bibliographystyle{aasjournal}
\bibliography{bibliography}

\begin{center}
\renewcommand\tabcolsep{4.pt}
\begin{longtable*}{c c c c c c c c c c c c c c}
\caption{Detailed information of the coronal microjets}
\label{paras} \\
\hline
Jet&Date&ST&Duration&Location&Speed&Length&Width&\lya&AIA&OP&Lane&Log (Ne) \\
  ID  &   2020-     &[UT] &[s]&X/Y& [\kms]  &  [Mm] &[Mm]    &&&&&[cm$^{-3}$]       \\\hline
01  & 05-20 & 22:03:08 & $>$530 & 932/238  & 46  & / & / &  / & yes & yes & yes & 9.15 \\
02  & 05-20 & 21:15:48 & 370 & 1693/420  & 46 & 5.3  & 0.7 & / & yes  & yes & yes & 9.15 \\
03  & 05-21 & 16:10:38 & $>$305 &  618/386  &  69 & 10.0 & 1.2 & / &  weak & yes & yes & /  \\
04  & 05-21 & 16:10:08 & 60 & 1483/430  & 52  & 4.0 & 0.4 & /  & yes & yes & yes & 9.35 \\
05  & 05-21 & 16:42:38 & 180 & 1056/381  & 41  & 2.7 & 0.5 & / & yes & yes & yes & 9.15 \\
06  & 05-21 & 16:40:58 & 50 &  1643/97 & 50  & 2.0 & 0.4 & /  & yes & yes & yes & 9.17 \\
07  & 05-21 & 16:43:58 & 190 &  1931/687 & 111  & 7.9 & 1.0 & /  & yes & maybe & yes & 9.03 \\
08  & 05-21 & 17:03:13 & 165 & 1403/477 & 71  & 4.2 & 0.5 & /  & yes & yes & close & 9.15 \\
09  & 05-21 & 17:11:58 & 300 &  1371/187 & 29  & 4.2 & 1.0 & /  & yes & maybe & yes & 9.03 \\
10  & 05-21 & 16:45:38 & 80 &  1327/1356 & 75  & 4.8 & 0.4 & /  & yes & no & no & 9.20 \\
11  & 05-21 & 16:42:58 & 230 & 1138/1595 & 179  & 9.4 & 0.9 & /  & weak  & yes & yes & / \\
12  & 05-21 & 17:20:58 & 180 & 169/1081 & 30  & 8.1 & 0.6 & /  & yes & yes & yes & 9.26 \\
13  & 05-21 & 16:59:58 & 210 & 1921/1368  & 59  & 6.5 & 1.0 & /  & weak & yes & close & / \\
14  & 05-30 & 14:50:58 & 80 &  678/1052 & 123  & 5.4 & 0.7 & yes  & yes & maybe & yes & 9.15 \\
15  & 05-30 & 14:51:38 & 45 & 1172/1544 & 72  & 2.8 & 0.5 & no  & no & no & yes & / \\
16  & 05-30 & 14:49:22 & $>$96 & 1318/1934  & 43  & 2.9 & 0.6 & no  & yes & maybe & yes & 9.24 \\
17  & 10-19 & 20:02:48 & 564 &  1095/417 & 70  & 12.4 & 1.1 & /  & / & / & / & / \\
18  & 10-19 & 20:21:12 & 300 & 1118/662  & 43  & 11.7 & 2.9 & /  & / & / & / & / \\
19  & 10-19 & 20:15:12 & 300 &  1034/757 & 84  & 5.7 & 1.0 & /  & / & / & / & / \\
20  & 10-19 & 20:00:41 & 199 & 1996/237  & 130  & 8.6 & 1.2 & /  & / & / & / & / \\
21  & 10-19 & 19:59:36 & 576 &  1219/929 & 41  & 7.5 & 1.6 & /  & / & / & / & / \\
22  & 10-19 & 20:22:36 & 84 & 1859/1056  & 132  & 6.5 & 0.7 & / & / & / & / & / \\
23  & 10-19 & 20:28:00 & $>$204 &  1519/1601 & 87  & 5.4 & 1.0 & /  & / & / & / & / \\
24  & 10-22 & 14:25:49 & 160 &  1556/1021 & 69  & 3.4 & 0.8 &  / & / & / & / & / \\
25  & 10-22 & 14:27:39 & $>$540 & 796/1984  & 77  & 9.5 & 2.7 &  / & / & / & / & / \\
26  & 11-19 & 11:55:35 & 225 & 209/138  & 65  & 5.7 & 0.8 & no  & / & / & yes & / \\
27  & 11-19 & 13:03:21 & 540 &  500/191 & 74  & 7.8 & 1.3 & no  & / & / & yes & / \\
28  & 11-19 & 14:07:51 & 135 &  131/261 & 77  & 6.1 & 0.8 & /  & / & / & / & / \\
29  & 11-19 & 12:19:51 & 420 &  714/419 & 36  & 7.4 & 1.6 & no  & / & / & yes & / \\
30  & 11-19 & 13:19:06 & 270 & 755/641  & 61  & 4.7 & 1.1 &  yes & / & / & yes & / \\
31  & 11-19 & 12:10:21 & 225 &  1068/181 & 56  & 6.3 & 0.9 & no  & / & / & yes & / \\
32  & 11-19 & 13:53:36 & 615 &  1240/216 & 48  & 20.0 & 1.7 & yes  & / & / & yes & / \\
33  & 11-19 & 13:58:51 & 405 &  1284/178 & 60  & 4.1 & 0.8 & no  & / & / & yes & / \\
34  & 11-19 & 12:05:20 & 271 &  1038/470 &  45 & 5.2 & 1.3 & yes  & / & / & yes & / \\
35  & 11-19 & 14:23:51 & $>$495 &  1252/519 & 18  & 7.2 & 0.8 & no  & / & / & yes & / \\
36  & 11-19 & 13:54:06 & 455 &  290/1178  &  24 & 6.5 & 0.8 & yes  & / & / & yes & / \\
37  & 11-19 & 14:19:51 & 150 &  885/1167 & 54  & 5.2 & 1.1 & yes  & / & / & yes & / \\
38  & 11-19 & 13:29:36 & 150 &  1145/1064 &  106 & 5.2 & 0.8 & no  & / & / & yes & / \\
39  & 11-19 & 13:57:51 & 180 & 1115/1026  & 65  & 6.6 & 0.9 & no  & / & / & yes & / \\
40  & 11-19 & 12:05:20 & 256 & 1658/952  & 39  & 4.1 & 0.9 & yes  & / & / & yes & / \\
41  & 11-19 & 12:20:36 & 255 & 1939/992  & 42  & 6.2 & 1.0 &  no & / & / & yes & / \\
42  & 11-19 & 14:18:21 & 150 &  82/1375 & 38  & 6.7 & 0.7 &  / & / & / & / & / \\
43  & 11-19 & 12:55:06 & $>$2085 & 436/1330  & 30  & 23.4 & 1.2 &  no & / & / & no & / \\
44  & 11-19 & 13:52:21 & 330 & 923/1519  & 41  & 7.3 & 1.5 & no  & / & / & yes & / \\
45  & 11-19 & 13:18:06 & 315 & 1167/1362  & 44  & 9.8 & 1.4 & yes  & / & / & yes & / \\
46  & 11-19 & 14:04:51 & 675 &  1284/1459 & 96  & 15.6 & 1.2 & yes  & / & / & yes & / \\
47  & 11-19 & 13:01:51 & 270 &  386/1867  & 34  & 13.5 & 1.1 & yes  & / & / & yes & / \\
48  & 11-19 & 13:55:36 & 240 & 140/1901  & 27  & 13.3 & 1.8 &  / & / & / & / & / \\
49  & 11-19 & 13:16:06 & 525 & 831/1934  & 38  & 10.3 & 1.5 &  no & / & / & yes & / \\
50  & 11-19 & 14:07:51 & 465 & 1462/1774  & 47  & 9.2 & 1.0 & yes  & / & / & yes & / \\
51  & 11-19 & 11:54:50 & 225 & 1077/1267  &  85 & 15.5 & 1.2 & no  & / & / & yes & / \\
52  & 11-19 & 12:30:36 & $>$90 &  1102/177 & 63  & 8.4 & 0.8 & no  & / & / & yes & / \\
\hline
\end{longtable*}
\tablenotetext{}{\textbf{ST:} the starting time of one microjet on the Sun, inferred from the HRI$_{EUV}$ images. \textbf{Location:} the location of a microjet in the pixel coordinate of a HRI$_{EUV}$ image. \textbf{\lya:} if a microjet reveals a signature in the HRI$_{Ly\alpha}$ images. \textbf{AIA:} if a microjet reveals a signature in AIA EUV images. \textbf{OP:} if a microjet is associated with opposite magnetic polarities. \textbf{Lane:} if a microjet originates from the network lane. \textbf{Jet-01} is a complex event, including several microjet-like structures that appear simultaneously at two adjacent locations. Therefore, it is difficult to estimate a length and a width. The projected speed was calculated from the AIA 171 \AA\ images.}
\end{center}

\end{document}